\documentclass[prl, showpacs, twocolumn, floatfix]{revtex4}

\usepackage{graphicx}
\usepackage{amsmath, amsfonts, amssymb, bm}

\begin{document}

\title{Quantum correlations of an atomic
ensemble via a classical bath}

\author{Mihai~\surname{Macovei}$\dagger$}
\email{macovei@usm.md}

\author{J\"org~\surname{Evers}} 
\email{joerg.evers@mpi-hd.mpg.de}

\author{Christoph~H.~\surname{Keitel}}
\email{keitel@mpi-hd.mpg.de}

\affiliation{Max-Planck-Institut f\"ur Kernphysik, 
 Saupfercheckweg 1, D-69117 Heidelberg, Germany}
\date{\today}

\begin{abstract}
Somewhat surprisingly, quantum features can be extracted 
from a classical bath. 
For this, we discuss a sample of three-level atoms in ladder 
configuration interacting only via the surrounding bath, and show that the
fluorescence light emitted by this system exhibits non-classical properties.
Typical realizations for such an environment are thermal baths for microwave 
transition frequencies, or incoherent broadband fields for optical transitions.
In a small sample of atoms, the emitted light
can be switched from sub- to super-poissonian and from 
anti-bunching to super-bunching controlled by 
the mean number of atoms in the sample.
Larger samples allow to  generate super-bunched light over a wide 
range of bath parameters and thus fluorescence light intensities.
We also identify parameter ranges where the fields emitted on the two 
transitions are correlated or anti-correlated, such that the 
Cauchy-Schwarz inequality is violated. As in a moderately strong  baths
this violation occurs also for larger numbers of atoms, 
such samples exhibit mesoscopic quantum effects. 
\end{abstract}

\maketitle

Initiated by Dicke~\cite{dik_et}, 
ensembles of few-level emitters interacting collectively with an 
environmental reservoir have
been shown to be a source for many remarkable effects and 
applications~\cite{dik_et,pr_et,has,THexp,state,ag_et,bog,mek1,EnM,kn_et}.
In the recent past, this interest was renewed by the possible 
applications of such samples to quantum communications and 
logic devices. For example, the production of non-classical
light has been a subject of several recent experiments~\cite{rempe,exp}.
On the other hand, it was demonstrated that long-time
entanglement between two arbitrary qubits can be generated if they 
interact with a common  bath~\cite{ent1}. Thus, 
quantum features can be induced through the interactions with a classical
electromagnetic field. This is not obvious, as, usually, it is believed 
that an interaction with a large environmental reservoir leads to 
decoherence. 
Thermal light may also produce effects like ghost imaging 
or sub-wavelength interference~\cite{therm}, which otherwise
are known to occur for entangled light~\cite{ent}. 
These results are of especial interest, as non-classical driving fields
are often hard to produce experimentally with adequate intensities.
Therefore, schemes which allow to extract quantum features from atoms 
in an otherwise classical external setup are  highly desirable.

In this Letter, we demonstrate that such a conversion scheme may be
implemented with an ensemble of atomic few-level systems subject 
to a classical bath. For this, we discuss a three-level 
setup in ladder configuration, and show that the fluorescence
light emitted spontaneously on the two transitions has
non-classical properties. The bath could e.g. be a thermal
bath for atoms with microwave transition frequencies, or incoherent
broadband driving for the optical frequency region.
In particular, we demonstrate that in a small sample of atoms,
the mean number of atoms in the sample allows to switch the
light emitted on one of the transitions from sub- to super-poissonian
statistics, and from anti- to super-bunching. 
Here, both super- and anti-bunched light with super-poissonian 
statistics can be produced.
Larger samples, on the other hand, can be used
to generate super-bunched light over a wide range of bath parameters, 
thus enabling a control of the intensity of the strongly correlated light.
We also identify parameter ranges where the fields emitted on the two 
transitions are correlated or anti-correlated.
As an application for this, we show that the 
Cauchy-Schwarz inequalities are violated for a moderately strong reservoir 
and large samples, thus demonstrating
a mesoscopic quantum effect. 

\begin{figure}[b]
\includegraphics[height=2.7cm]{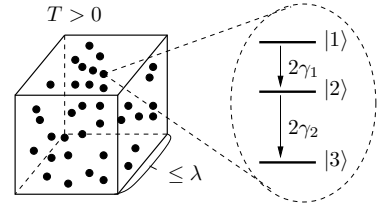}
\caption{\label{fig-system}The system setup. A sample of $N$ atoms, each 
with three atomic states in ladder configuration, is confined to a region
small as compared to the emission wavelengths $\lambda$. 
The whole setup is placed in a classical bath.
The excited states decay spontaneously with rates $2\gamma_{1(2)}$,
giving rise to fluorescence lights with non-classical features.}
\end{figure}


The basic element of our study is a sample of $N$ identical non-overlapping
three-level radiators which interact with a classical reservoir, see
Fig.~\ref{fig-system}. The
emitters are located within a volume with linear dimensions smaller than
the relevant emission wavelengths $\lambda _{12},\lambda _{23}$, and
densities low enough to avoid collisions 
(Dicke model)~\cite{dik_et,pr_et,has,THexp,state,ag_et,bog,mek1,EnM}.
The excited atomic level $|1\rangle $ ($|2\rangle $) spontaneously
decays to the state $|2\rangle $ ($|3\rangle $) with a decay rate 
$2\gamma _{1}$ ($2\gamma _{2}$). The only external driving is via 
the surrounding bath, which induces transitions among the atomic levels.
In the microwave region, the bath could be a thermal bath, where
the rates are proportional to the mean thermal photon number at the
corresponding transition frequencies. In the optical frequency region,
thermal excitations are negligible, and the bath can be realized by a
pseudo-thermal bath, i.e. broadband incoherent driving fields perpendicular
to the observation direction. Then,
the rates depend on the field strength of the incoherent fields.
In the usual mean-field, Born-Markov and rotating-wave approximations, the 
system is described by the following master 
equation~\cite{ag_et,mek1}: 
\begin{eqnarray}
\dot{\rho}(t)&=& -\gamma _{1}(1+\bar{n}_{1})[S_{12},S_{21}\rho] 
     - \gamma _{2}(1+\bar{n}_{2})[S_{23},S_{32}\rho ] \nonumber \\
&-& \gamma _{1} \bar{n}_{1}[S_{21},S_{12} \rho ] 
     - \gamma _{2}\bar{n}_{2}[S_{32},S_{23}\rho ]+\textrm{ h.c.}\,. 
      \label{master}
\end{eqnarray}
Here, an overdot denotes differentiation with respect to time. For thermal
baths, $\bar{n}_{i }=[\exp (\beta \hbar 
\omega _{i,i+1})-1]^{-1}$ is the 
mean thermal photon number at transition frequency 
$\omega _{i,i+1} = \omega _{i}-\omega _{i+1}$ and 
for temperature $T$, where $\beta =(k_{B}T)^{-1}$ with $k_{B}$ as the 
Boltzmann constant. For pseudo-thermal baths, 
$\bar{n}_i = R_{i,i+1} d^{2}_{i,i+1}/(\gamma_{i}\hbar^{2})$,
where $R_{i,i+1}$ describes the strength of the incoherent 
pumping~\cite{pseudo}.
It is important to note that Eq.~(\ref{master}) contains
collective atomic operators 
$S_{i j }  = \sum_{k=1}^{N}|i \rangle_{k}{}_{k}\langle j |$, which
describe populations for $i=j$, transitions for
$i \not=j $, and which obey the commutation relation 
$[S_{i  j },S_{i^{\prime}j^{\prime }}] = 
 \delta _{j i ^{\prime}}S_{i j ^{\prime }} 
 - \delta _{j ^{\prime }i }S_{i ^{\prime }j }$
 ($i,j \in \{1,2,3\}$)~\cite{mek1}.

The steady-state limit of Eq.~(\ref{master}) can conveniently be evaluated 
with the help of coherent atomic states $|N,n,m\rangle$ for the su(3) 
algebra, which
 denote a symmetric collective state of $N$ atoms with $n$ atoms in 
bare state $|1\rangle$, $m-n$ in bare state $|2\rangle$,
and $N-m$ atoms in bare state $|3\rangle$ with
$0\leq n\leq N$, $n\leq m\leq N$~\cite{ag_et,bog,mek1}.
The diagonal elements $P_{nm} = \langle N,n,m|\rho_{ss}|N,n,m\rangle$ of 
the steady-state density operator $\rho_{ss}$ evaluate to:
\begin{eqnarray}
P_{nm} &=& ( 1 - \eta_{2} )\eta^{n}_{1}\eta^{m}_{2} \nonumber\\
&\times& \biggl [\frac{1-(\eta_{1}\eta_{2})^{N+1}}{1-\eta_{1}\eta_{2}} 
- \eta^{N+1}_{2}\frac{1-\eta^{N+1}_{1}}{1-\eta_{1}}\biggr ]^{-1}\,, 
 \label{s_sol}
\end{eqnarray}
with $\eta_{i} = \bar n_{i}/(1+\bar n_{i})$, $(i \in \{1,2\})$. From
Eq.~(\ref{s_sol}), the atomic expectation values 
can easily be evaluated.


We now turn to our main interests in this study, the coherence 
properties of the collective fluorescent light generated on the transitions
$|1\rangle \to |2\rangle$ and $|2\rangle \to |3\rangle$.
The photons emitted on the two transitions are distinguishable by their
polarizations and frequencies, and can be detected e.g. by a pair of
single-photon detectors or by atomic state detection~\cite{THexp,state}.
The second-order coherence function is defined 
as~\cite{qc1,qc2}:
\begin{eqnarray}
g^{(2)}_{ij}(\tau) = \frac{\langle J^{+}_{i}(t)J^{+}_{j}(t+\tau)
J_{j}(t+\tau)J_{i}(t)\rangle}
{\langle J^{+}_{i}(t)J_{i}(t) \rangle 
\langle J^{+}_{j}(t)J_{j}(t) \rangle} \,,  \label{g2}
\end{eqnarray}
where $i,j\in\{1,2\}$ with $J_{1}=S_{21}$ and $J_{2}=S_{32}$. The 
quantity $g^{(2)}_{ij}(\tau)$  can be interpreted as a measure for 
the probability for detecting one photon emitted on transition
$i$ and another photon emitted on transition $j$ with time delay $\tau$.
$g^{(2)}_{ij}(0)<1$ characterizes sub-poissonian,
$g^{(2)}_{ij}(0)>1$ super-poissonian, and $g^{(2)}_{ij}(0)=1$ poissonian 
photon statistics. $g^{(2)}_{ij}(\tau) > g^{(2)}_{ij}(0)$ is the condition for
photon anti-bunching, whereas $g^{(2)}_{ij}(\tau) < g^{(2)}_{ij}(0)$
means bunching. We further define super-bunching as bunching with
$g^{(2)}_{ij}(0)>2$~\cite{anti,sub}. More specific, correlation
functions with $i=j$ describe the photon statistics of the fluorescence
light emitted on a single atomic transition, and $g^{(2)}_{i \not = j}(0)$
the cross-correlations between the photon emission on two different 
transitions. 
%
\begin{figure}[b]
\includegraphics[height=4cm]{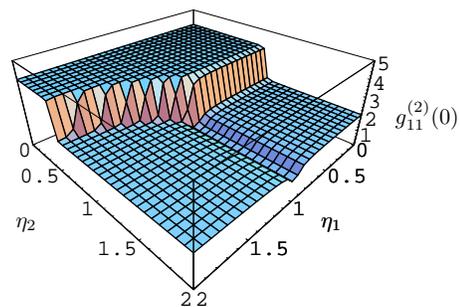}
\caption{\label{fig-g11}Plot of the second-order correlation function 
$g^{(2)}_{11}(0)$ against bath parameters $\eta_{1},\eta_{2}$.
The number of atoms in the sample is $N=150$.}
\end{figure}
%
%
We also need to consider the fluorescence intensities
$G^{(1)}_{i}(0) = \langle J^{+}_{i}J_{i} \rangle_s$
of the two transitions.
For example, applications may require particularly strong or 
weak non-classical fields. On the other hands,
in the microwave region, the signal from the sample competes with noise
from the surrounding heat bath proportional to $\bar{n}_i$, which 
especially for small samples
of atoms with low signal can render experimental verifications difficult.
For bath parameters
$0<\eta_{1},\eta_{2} <1$ and larger samples $N\gg 1$ such that
$\eta^{N}_{1(2)} \to 0$,  one has 
$G^{(1)}_{1}(0) \approx \bar n_{1}\bar n_{2}/[1+\bar n_{1}/(1+\bar n_{2})]$,
$G^{(1)}_{2}(0) \approx \bar n_{2}N$.
Thus $G^{(1)}_{1}(0)$ does not depend on $N$ explicitly, 
while $G^{(1)}_{2}(0)$ 
increases linearly with $N$. 
In the strong field limit 
$(\eta_{1},\eta_{2} \to 1)$, one has
$G^{(1)}_{1(2)}(0) = N(3+N)/12 \sim N^2$.

\begin{figure}[t]
\includegraphics[width=4.2cm]{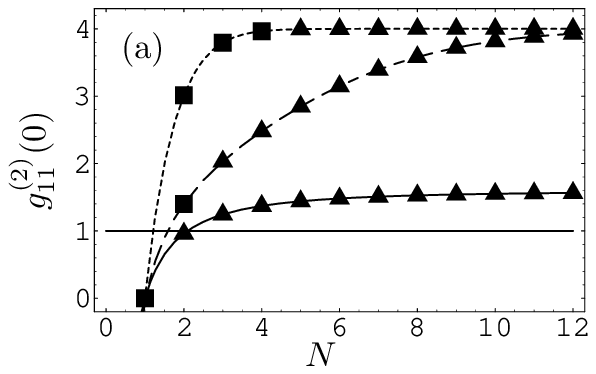}
\includegraphics[width=4.2cm]{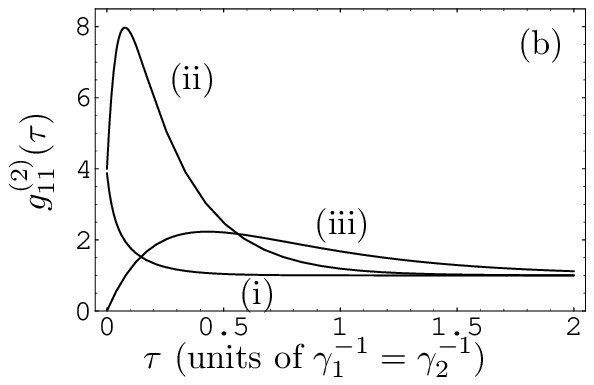}
\caption{\label{fig-smallsample}(a) The second-order correlation
function $g^{(2)}_{11}(0)$ for small numbers of atoms $N$ in the
sample. Solid, long dashed and short dashed curves are for 
$\eta_{1} = \eta_{2} = 1, 0.5$ and $0.1$, respectively. Non-integer
values for $N$ are possible if $N$ is a mean number of atoms, e.g.
for atoms crossing a cavity field. Triangles (squares) denote 
bunching (anti-bunching). 
(b) $g^{(2)}_{11}(\tau)$ versus delay time $\tau$.
(i) $N=15, \eta_1=\eta_2=0.6$, (ii) $N=6, \eta_1=0.8, \eta_2=0.05$,
(iii) $N=1, \eta_1=\eta_2=0.2$.}
\end{figure}

\begin{figure}[t]
\includegraphics[width=8cm]{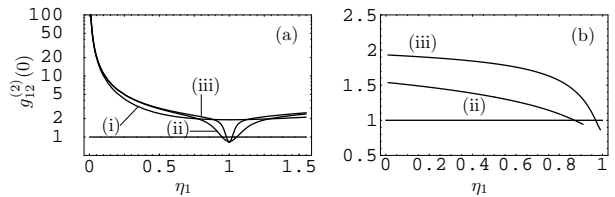}
\caption{\label{fig-g12} The  cross-correlation function
 $g^{(2)}_{12}(0)$ against bath parameters
 $\eta_{1}=\eta_{2}$. The number of atoms in the sample is
 (i) $N=2$, (ii) $N=50$, and (iii) $N=150$. 
 (a) $\eta_1=\eta_2$,
 (b) $\eta_2$ chosen such that the SNR on both transitions
 is above 10. }
\end{figure}

The first results are shown in Fig.~\ref{fig-g11}, where the correlation 
function $g^{(2)}_{11}(0)$ of the light emitted on transition 
$|1\rangle \to |2\rangle$ is plotted against $\eta_1$ and $\eta_2$ for a 
sample of $N=150$ atoms. It can be seen that photons with super-poissonian
statistics are generated on this transition for moderate  
baths. In this region, the emitted light is super-bunched except
for small values of $\eta_2$, where the light is anti-bunched.
This range of $\eta_2$ for anti-bunching depends on the number of atoms 
$N$ and $\eta_1$. In Fig.~\ref{fig-g11}, for $\eta_1<0.6$, anti-bunching
requires $\eta_2<0.01$. For smaller samples, however, the range 
for anti-bunching increases.
If $\eta_2$ is small, then almost all atoms are in the ground state 
$|3\rangle$  due to collective effects, and on average at most one atom gets 
excited to $|1\rangle$ 
to emit light contributing to $g^{(2)}_{11}(0)$. Thus the light is
anti-bunched. For higher $\eta_2$, more atoms can be excited to $|1\rangle$
simultaneously, and the light is bunched.
%
As the super-bunched photons are produced over a wide range 
of values for $\eta_i$, the intensity of the generated light can be 
controlled via the bath parameters $\bar n_{1}, \bar n_{2}$ 
from very weak up to intense flux.
In other words, by modifying the reservoir characteristics, we can obtain a 
low or an intense flux of strongly correlated photons. 
This setup is particularly suitable for microwave transitions, as then
thermal baths with high values for $\eta_i$ can be achieved, and larger
samples of atoms allow for a good signal-to-noise ratio (SNR). 
An optical realization, however, is also possible.
We now turn to the ``channel'' around $\eta_{1}=1, \eta_{2} \ge 1$ in 
Fig.~\ref{fig-g11}. For $\eta_1=\eta_2=1$, one finds 
$g^{(2)}_{11}(0) = 8(N-1)(N+4)/[5N(N+3)]$, which for $N \to \infty$ goes 
to $8/5$.
The corresponding limit for $\eta_{1}=1, \eta_{2} > 1$ yields $6/5$.
%
This ``channel''
can be understood by noting that for these parameters the sample
acts collectively, i.e. $G^{(1)}_{i}(0) \propto N^{2}$, $(i\in\{1,2\})$, 
resulting  in a close to coherent photon statistics that corresponds
to a superfluorescent atomic sample~\cite{dik_et,pr_et,has}.
It should be emphasized here that while a thermal
reservoir or a direct incoherent pumping of the transitions
only admits for values $0< \eta_{1}, \eta_{2} \le 1$, we have 
also used larger values for these parameters in 
Figs.~\ref{fig-g11},\ref{fig-g12}.  
Such situations may occur if additional driving fields are applied to the 
sample of atoms, e.g. an incoherent repumping from the lower to the upper 
atomic states~\cite{mek1}. We stress, however, that our main results are
obtained without such driving.

In the following, we discuss the special case of a small collection 
of atoms. This is of particular interest for an experimental
verification  in the optical region, whereas in the microwave region 
the small number of atoms makes it hard to obtain a decent SNR 
against the thermal background.
In Fig.~\ref{fig-smallsample}(a), we show  
$g^{(2)}_{11}(0)$ against the number of atoms in the sample for different 
values of $\eta_1 = \eta_2$. Consider, for example, an experimental setup 
with an atomic beam passing through a low quality cavity.
Then, depending on the mean number of atoms which are simultaneously inside 
the cavity and on the bath parameters, 
switching between sub- and super-poissonian statistics or 
anti-bunching and super-bunching of the emitted light 
can be observed~\cite{rempe}. 
In the figure, squares (triangles) denote anti-bunching (bunching)
of the emitted photons. Note that
together with a super-poissonian statistics, both bunching and anti-bunching
can be observed. 
Only collective states $|N,1,m\rangle$ with a single atom in
state $|1\rangle$ may lead to anti-bunching. All other states
$|N,n\!>\!1,m\rangle$ contribute to bunching. The total system
behavior depends on the ratio of these two contributions.
With increasing bath strength and increasing number of atoms, 
more bunching states $|N,n\!>\!1,m\rangle$ are available and populated, 
such that the switching from anti-bunching to super-bunching occurs.
%
%
Some examples for $g^{(2)}_{11}(\tau)$ versus delay time $\tau$ 
are shown in Fig.~\ref{fig-smallsample}(b). Curve (i) shows
super-bunching for $N=15$ and $\eta_1=\eta_2=0.6$. Starting
from an initial value close to 4, the correlation function drops rapidly
to unity with increasing $\tau$. Example (ii) shows anti-bunching
with super-poissonian photon statistics and large intermediate values
of $g^{(2)}_{11}(\tau)$. The maximum value of the correlation function
can be further increased, however, at the cost of intensity.
As reference, (iii) shows an evolution for the single-atom case.

The cross-correlations $g^{(2)}_{i\not =j}(0)$ also show
non-classical behavior.
For an atomic sample in a weak bath,
$g^{(2)}_{12}(0)$ is much larger than unity as shown in Fig.~\ref{fig-g12},
 indicating super-poissonian light statistics, which is accompanied
 by strong correlation between 
the fluorescence light radiated on both atomic transitions,
i.e. cross super-bunching. 
The reason is that then atoms which decay
from $|1\rangle$ to $|2\rangle$ also decay further to $|3\rangle$
with a high probability.
For stronger baths, however, larger samples exhibit bunched 
sub-poissonian light.
Then $\lim_{\{\eta_{1},\eta_{2}\}\to 1}g^{(2)}_{12}(0) 
 = 4(N+2)(N+4)/[5N(N+3)]$, with limit $4/5<1$ for $N \to \infty$. 
In this case, atoms decaying from $|1\rangle$ to $|2\rangle$ are
repumped by the bath rather than decaying further to $|3\rangle$.

The other cross-correlation, $g^{(2)}_{21}(0)$, is below 
unity for a weak  bath,
as a transition $|2\rangle$ to
$|3\rangle$ cannot directly be followed by a transition $|1\rangle$
to $|2\rangle$ without an extra excitation. 
%
%
For small samples $(N<8$), or for medium samples ($N\sim 40$) with strong
bath on transition 1 ($\eta_1 \to 1$), the emitted light is
anti-bunched, thus showing collective cross anti-bunching. If both
transitions are driven strongly, one finds cross anti-bunching with 
$G^{(1)}_{i}(0) \propto N^{2}$.
For larger samples and smaller $\eta_1$, the light is anti-bunched
for low values of $\eta_2$, but switches to bunched light with 
increasing $\eta_2$.

\begin{figure}[t]
\includegraphics[width=6.5cm,height=3.5cm]{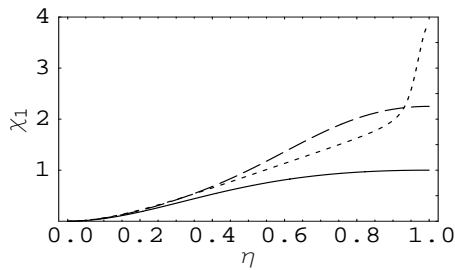}
\caption{\label{fig-csv} The Cauchy-Schwarz parameters $\chi_{1}$ 
shown versus bath parameters 
$\eta_{1}=\eta_{2} \equiv \eta$. The solid, long-dashed 
and short-dashed curves are plotted for $N=4, 10$ and $150$.}
\end{figure}

As an application for the non-classical features, we now show that
the light emitted from the sample of atoms violates the Cauchy-Schwarz 
inequalities (CSI) \cite{csv}. The 
CSI are violated if 
$\chi_{1(2)} = g^{(2)}_{11}(0)g^{(2)}_{22}(0)
 / [g^{(2)}_{12(21)}(0)]^{2}<1$,
i.e., if the cross-correlations between photons emitted on two different
transitions are larger than the correlation between photons emitted from 
the individual levels. 
Fig.~\ref{fig-csv} shows the violation of the CSI function for moderately strong 
baths. Within the Dicke model, this violation is present 
for any number of atoms in the sample, thus demonstrating a 
mesoscopic quantum effect.
In addition, 
$\chi_{1}$ is always smaller than unity for $N \le 3$ and the entire 
range of $0 \le \eta_{1},\eta_{2} \le 1$
($\lim_{\{\eta_{1},\eta_{2}\} \to 1}\chi_{1} = 4[(N-1)/(N+2)]^{2}$),
while $\chi_{2}$ is larger than unity for any 
number of atoms and for any values of $\eta_{1},\eta_{2}$.
The CSI violation can best be observed in the optical region,
as low values of $\eta_i$ are favorable, and as it is more difficult
to obtain a decent SNR in the microwave region.

In summary, we have demonstrated quantum features in the fluorescence
light of a sample of atoms driven only by a  surrounding classical 
bath. We discussed both thermal baths with microwave atomic transition 
frequencies and pseudo-thermal baths for optical transition 
frequencies as realizations of the bath.
For small samples, a change of the mean number
of atoms in the sample induces sensitive switching between 
sub- or super-poissonian statistics and anti-bunching 
or super-bunching of the light emitted on one of the transitions.
For appropriate bath parameters, even mesoscopic samples exhibit 
anti-bunching.
As an application, we have shown that the Cauchy-Schwarz inequalities
are violated in our system over a wide range of parameters.

{\small $^\dagger$ Permanent address: \it{Technical University of 
Moldova, Physics Department, \c{S}tefan Cel Mare Av. 168, MD-2004
Chi\c{s}in\u{a}u, Moldova.}}


\end{document}